\documentclass[preprint,12pt]{elsarticle}

\usepackage{amssymb}

\usepackage{color}

\journal{Applied Mathematics and Computation - accepted for publication - }

\begin{document}

\begin{frontmatter}




\title{Isolation effects in a system of two mutually communicating identical patches}


\author[unifal]{D. J. Pamplona da Silva}
\ead{pamplonasa@gmail.com}
\author[unifal]{R. P. Villar}
\author[unifal]{L. C. Ramos}

\address[unifal]{Universidade Federal de Alfenas - UNIFAL-MG, Rodovia BR 267 - km 533 - 37701-970
Po\c{c}os de Caldas, Brazil}

\begin{abstract}
Starting from the Fisher-Kolmogorov-Petrovskii-Piskunov equation (FKPP) we model the dynamic of a diffusive system with two mutually communicating identical patches and isolated of the remaining matrix. For this system we find the minimal size of each fragment in the explicit form and compare with the explicit results for similar problems found in the literature. From this comparison emerges an unexpected result that for a same set of the parameters, the isolated system studied in this work with size $L$, can be better or worst than the non isolated systems with the same size $L$, uniquely depending on the parameter $a_{0}$ (internal conditions of the patches). Due to the fact that this result is unexpected we propose an experimental verification.

\end{abstract}


\end{frontmatter}


\section{Introduction}
\label{Introduction}

In the study of population dynamics, it is used many tools like me\-ta\-po\-pu\-la\-tions \cite{Johst,Snall}, diffusive systems \cite{Argolo, Ducrot}, with one \cite{Skellam} and more \cite{Das} species interacting in many forms \cite{Amarasekare,Beraldo,Nathaniel,Singh}.

The problem, of a single species moving in a diffusive pattern is largely \cite{Azevedo,Fisher,Kenkre2011,Lin,Nelson} modeled in literature by the equation of Fisher-Kolmogorov-Petrovskii-Piskunov (FKPP), that in one dimension is given by \cite{Artiles,Kenkre2008,Kenkre2003,Pamplona,Skellam}:
\begin{equation} \label{FKPP}
\frac{\partial \Phi}{\partial t} = D\frac{\partial^{2}\Phi}{\partial x^2}+a(x)\Phi-b\Phi^2,
\end{equation}
where $\Phi=\Phi(x,t)$ is the population density, $t$ is the time, $x$ is the spatial variable, $D$ is the diffusion coefficient, $a(x)$ is the growth rate and $b$ is a saturation constant (related to the carrying capacity).

\begin{figure}[ht]
\begin{center}
\includegraphics[height=3cm]{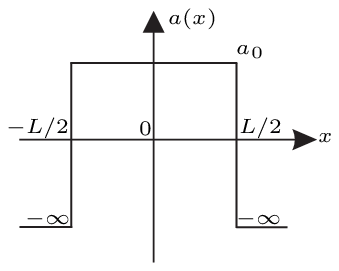}\hspace{0.5cm}
\includegraphics[height=3cm]{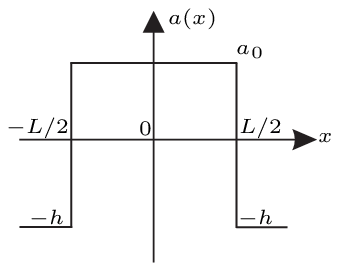}\hspace{0.5cm}
\includegraphics[height=2.7cm,width=4.5cm]{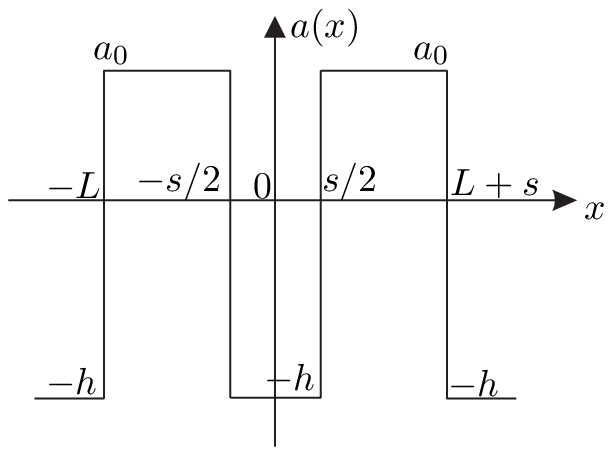}\\
a \hspace{4cm} b \hspace{4cm} c
\caption{\label{asdex} Representation of fragmented regions: {\bf a:} one patch isolated from the matrix {\bf b:} one patch non isolated from the matrix {\bf c:} Two identical patches immersed in a matrix, separated by a region of length $s$. In these three cases, the internal conditions of the patches are $a_{0}$, their lengths are $L$ and the life difficulty in the matrix is quantified by parameter $h$, except in the case of isolated systems, where life is impossible in the matrix.}
\end{center}
\end{figure}

\begin{figure}[h]
\includegraphics[width=4cm]{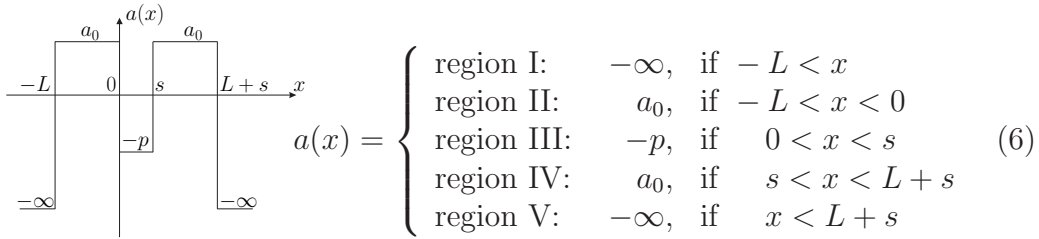}\nonumber
\caption{\label{a(x)d} Representation of an isolated system with two fragments mutually communicating. $L$ is the size of the patches, $a_{0}$ the internal growth rate and $p$ is the life difficulty level between the patches.}
\end{figure}

The function $a(x)$ is used to describe spatial heterogeneity, where we assume $a(x)>0$ as a life region, a zone good for life (patch, island, fragment). If $a(x)<0$, we assume as a death region, which is unfavorable for life. The profiles described in Figs. (\ref{asdex}) and (\ref{a(x)d}) represents examples of fragmented regions.

Using Ludwig arguments \cite{Ludwig}, reinforced in the literature \cite{Kenkre2008,Pamplona}, we consider the stead state of FKPP, Eq. (\ref{FKPP}) and neglected the nonlinear term $-\Phi^2$, to find the limit conditions between life region and death region. These con\-si\-de\-ra\-tions generate the equation:

\begin{equation} \label{FKPP-le}
D\frac{d^{2}\Phi}{d x^2}+a(x)\Phi=0.
\end{equation}

Many profiles of heterogeneity can be interesting to population dynamics because they represent real systems, but if the function $a(x)$ assumes strange forms, the solution of Eq. (\ref{FKPP-le}) can be difficult and unfeasible to find. One simple form of $a(x)$ interesting to the study of population dynamics is the piecewise constant function. In this case, we assume homogeneous regions where $a(x)>0$ like a patch and regions (homogeneous too) where $a(x)<0$ like the matrix or the separation between two neighbor patches such as those in Fig. (\ref{asdex}).

In the literature, it is possible to find profiles of $a(x)$ as piecewise constant function used to interpret biological growth systems. For example, there is the profile for one patch isolated of the matrix, Fig. (\ref{asdex}a), which the minimal size patch was found by Skellam \cite{Skellam} and confirmed by Kenkre \cite{Kenkre2003}, satisfying:
\begin{equation}\label{eqLsi}
L_{si}=\pi\sqrt{\frac{D}{a_{0}}}.
\end{equation}

Another example of one patch profile, but non isolated from the matrix, Fig. (\ref{asdex}b), was studied by Ludwig \cite{Ludwig} who presented an expression for the minimum size of the fragment, in the form:
\begin{equation}\label{eqLsn}
L_{sh}=2\sqrt{\frac{D}{a_{0}}} \arctan{\sqrt{\frac{h}{a_{0}}}}.
\end{equation}

There are studies for infinite numbers of patches \cite{Kenkre2008,Kraenkel}, but one interesting case that has an explicit form for minimal island size is the case of two identical fragments immersed in the matrix, Fig. (\ref{asdex}c), it was proposed by Kenkre \cite{Kenkre2008} who predicted Eq. (\ref{eqLdh}):
\begin{equation}\label{eqLdh}
L_{dh}=\sqrt{\frac{D}{a_{0}}}\left\{\arctan{\sqrt{\frac{h}{a_{0}}}}+\arctan{\left[\sqrt{\frac{h}{a_{0}}}\tanh{\left(\sqrt{\frac{h}{D}}\frac{s}{2}\right)}\right]}\right\}.
\end{equation}

In this article, we propose two identical patches isolated from the matrix, but mutually communicating, which is the main propose of this work. This profile is represented in Fig. (\ref{a(x)d}).

\begin{equation}\label{pdf}
\hspace{3.5cm}a(x)=\left \{\begin{array}{lrl}
\mbox{region I: } & -\infty,&\mbox{if}\,\, -L<x\\
\mbox{region II: } & a_{0},&\mbox{if}\,\, -L<x<0\\
\mbox{region III: } & -p,&\mbox{if}\quad\,\,\, 0<x<s\\
\mbox{region IV: } & a_{0},&\mbox{if}\quad\,\,\, s<x<L+s\\
\mbox{region V: } & -\infty,&\mbox{if}\quad\,\,\, x<L+s
\end{array}\right.
\end{equation}
\vspace{-4.0cm}

The solution to Eq. (\ref{FKPP-le}) with the profile addressed in Fig. (\ref{a(x)d}) and the continuity and boundary conditions return the explicit form of the minimum size $L_{di}$ of the patches in this case. Our concern with the calculation of the solution to the Eq. (\ref{FKPP-le}) is not great, then we do not present it in this work. Due to our phenomenological interest we stop our discussion at the relations of continuity and boundary conditions. This is like detailed in section \ref{Result}.

\section{Result}
\label{Result}

In order to solve Eq. (\ref{FKPP-le}) with the function $a(x)$ described by Eq. (\ref{pdf}), first we find  separately the solutions $\Phi_{I}(x)$ to $\Phi_{V}(x)$ in the regions from I to V respectively and after that we use the boundary and continuity conditions on the borders. We have:
\[
\Phi_{I}(x)=\Phi_{V}(x)=0,
\]
\[
\Phi_{II}(x)=A\cos{\mu_{n}x} + B \mbox{sen}{\mu_{n}x} \quad \mbox{ and }  \quad  \Phi_{IV}(x)=E\cos{\mu_{n}x} + F \mbox{sen}{\mu_{n}x},
\]
\[
\Phi_{III}(x)=Ce^{\nu_{n}x} + De^{-\nu_{n}x},
\]
where $\displaystyle \mu_{n}=\sqrt{\frac{a_{0}}{D}} \mbox{ and } \nu_{n}=\frac{p}{D}$.\\

{\noindent \bf Boundary and continuity conditions of $\Phi(x)$}
\begin{itemize}
\item In $x=-L: \Phi_{I}(-L)=\Phi_{II}(-L) \Rightarrow A=B \tan{\mu_{n}L}$, then:
\begin{equation}\label{Phi(II)}
\Phi_{II}(x)=B[\tan{\mu_{n}L}\cos{\mu_{n}x}+ \mbox{sen}{\mu_{n}x}].
\end{equation}

\item In $x=L+s$ from $\Phi_{IV}(L+s)=\Phi_{V}(L+s)$. This leads to:
\[
\Phi_{IV}(x)=F[\mbox{sen}{\mu_{n}x}-\tan{\mu_{n}(L+s)}\cos{\mu_{n}x}].
\]

\item In $x=0$ we have $\Phi_{II}(0)=\Phi_{III}(0)$, where using $\Phi_{II}$ from Eq. (\ref{Phi(II)}). Then,
\begin{equation}\label{eq8}
B\tan{\mu_{n}L}=(C+D).
\end{equation}

\item From $x=s$ where $\Phi_{III}(s)=\Phi_{IV}(s)$ we extract after some algebraic steps that involve identities related to the sum of angles:
\begin{equation}\label{eq10}
-F\sin{\mu_{n}L}=(Ce^{\nu_{n} s}+De^{-\nu_{n} s})\cos{\mu_{n}(L+s)}.
\end{equation}

\end{itemize}

\noindent{\bf Analogously, from $d\Phi(x)/dt$ continuity conditions}

\begin{itemize}

\item In $x=0$ immediately goes:
\begin{equation}\label{eq9}
B\mu_{n}=\nu_{n}(C-D).
\end{equation}
\item In $x=s$ we have, not immediately,
\begin{equation}\label{eq11}
F\mu_{n}\cos{\mu_{n}L}=(Ce^{\nu_{n} s}-De^{-\nu_{n} s})\nu\cos{\mu_{n}(L+s)}.
\end{equation}

\end{itemize}


By isolating $B$ at Eq. (\ref{eq8}), then substituting it in Eq. (\ref{eq9}) and dividing Eq. (\ref{eq10}) by Eq. (\ref{eq11}), we eliminate $B$ and $F$ and construct a system of equations to $C$ and $D$. On the matrix form the system reads:
\begin{equation}\label{apA12}
\left (\begin{array}{ll}
\displaystyle  \nu -\frac{\mu}{\tan{\mu L}} \quad& 
\displaystyle -\nu -\frac{\mu}{\tan{\mu L}} \\
\\
\displaystyle -\nu e^{\nu s}-\frac{\mu e^{\nu s}}{\tan{\mu L}} \quad& 
\displaystyle \nu e^{-\nu s}-\frac{\mu e^{-\nu s}}{\tan{\mu L}}\\
\end{array}\right)
\left (\begin{array}{l}
\displaystyle  C \\
\\ 
\displaystyle  D \\
\end{array}\right)
=
\left (\begin{array}{l}
\displaystyle  0 \\
\\ 
\displaystyle  0 \\
\end{array}\right).
\end{equation}

In order to find a non trivial solution to the system of Eq. (\ref{apA12}), we impose that the determinant is null. This requirement leads to the following equation:
\begin{equation}\label{apA13}
\left(\nu -\frac{\mu}{\tan{\mu L}}\right)^{2}e^{-\nu s}-\left(\nu +\frac{\mu}{\tan{\mu L}}\right)^{2}e^{\nu s}=0.
\end{equation}

By extracting the square root of Eq. (\ref{apA13}) and grouping the terms in $\mu$ and $\nu$ conveniently, we have:
\begin{equation}\label{apA15}
\nu(e^{-\nu s/2}-e^{\nu s/2})=\frac{\mu}{\tan{\mu L}}(e^{-\nu s/2}+e^{\nu s/2}).
\end{equation}

Once $\displaystyle\tanh{x}=\frac{e^{x}-e^{-x}}{e^{x}+e^{-x}}$, the Eq. (\ref{apA15}) assumes the compact form:
\begin{equation}\label{apA16}
\mu \cot{\mu L}=-\nu\tanh{\left(\nu\frac{s}{2}\right)},
\end{equation}
which leads to the following explicit function for $L(p,s)$:

\begin{equation}\label{apA17}
L(p,s)=\frac{1}{\mu}\left\{n\pi-\mbox{arccot}\left[\frac{\nu}{\mu}\tanh{\left(\nu\frac{s}{2}\right)}\right]\right\}, \mbox{ with } n\in \mathbb{Z}.
\end{equation}

The term $n\pi$ in Eq. (\ref{apA17}) indicates the periodic property of the function cotangent. In order to obtain the first positive determination, we choose $n=1$ and we use the identity $$\mbox{arccot }x=\frac{\pi}{2}-\arctan{x},$$ for comparison with the literature \cite{Kenkre2008,Ludwig}. The last equation enables us to rewrite Eq. (\ref{apA17}) in the following form,

\begin{equation}\label{eqLdi}
L_{di}=\sqrt{\frac{D}{a_{0}}}\left\{\frac{\pi}{2}+\arctan{\left[\sqrt{\frac{p}{a_{0}}}\tanh{\left(\sqrt{\frac{p}{D}}\frac{s}{2}\right)}\right]}\right\},
\end{equation}
which already contains our initial variables.

Eq. (\ref{eqLdi}), main analytical result of this article, is the explicit expression of minimal patch size of isolated system with two mutually comunicating identical patches. This result can be obtained from a previous paper \cite{Pamplona}, as a particular case, but not in an explicit form. Although the numerical agreement between both is perfect, the explicit form, presented here, enable us to explore the functional relation between the parameters.

\begin{figure}[ht]
\begin{center}
\includegraphics[height=9cm]{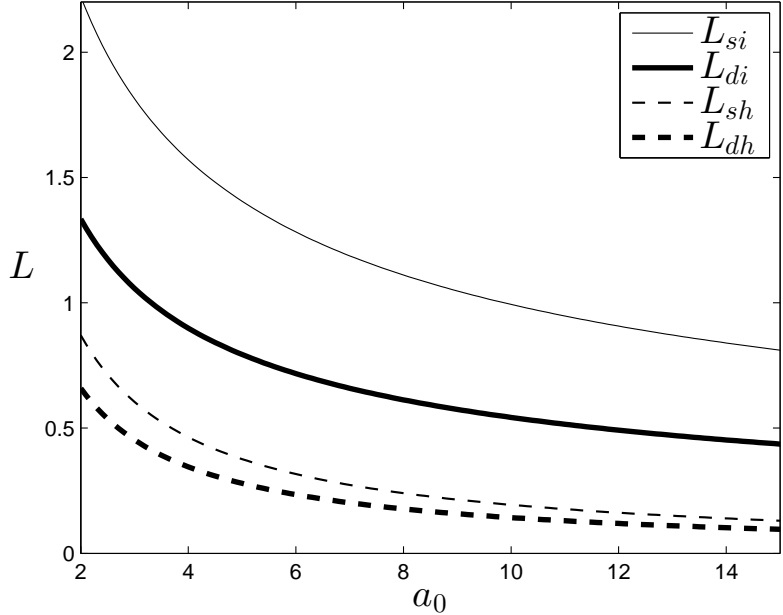}
\caption{\label{Lxa0_expected} Minimum size of fragments versus growth rate ($a _{0}$) for systems: isolated with one ($L_{si}$) or two ($L_{di}$) patches and one ($L_{sh}$) or two ($L_{dh}$) patches immersed in the matrix. All curves were plotted with parameters $h=1$, $p=1$, $s=1$, $D=1$.}
\end{center}
\end{figure}

\begin{figure}[ht]
\begin{center}
\includegraphics[height=9cm]{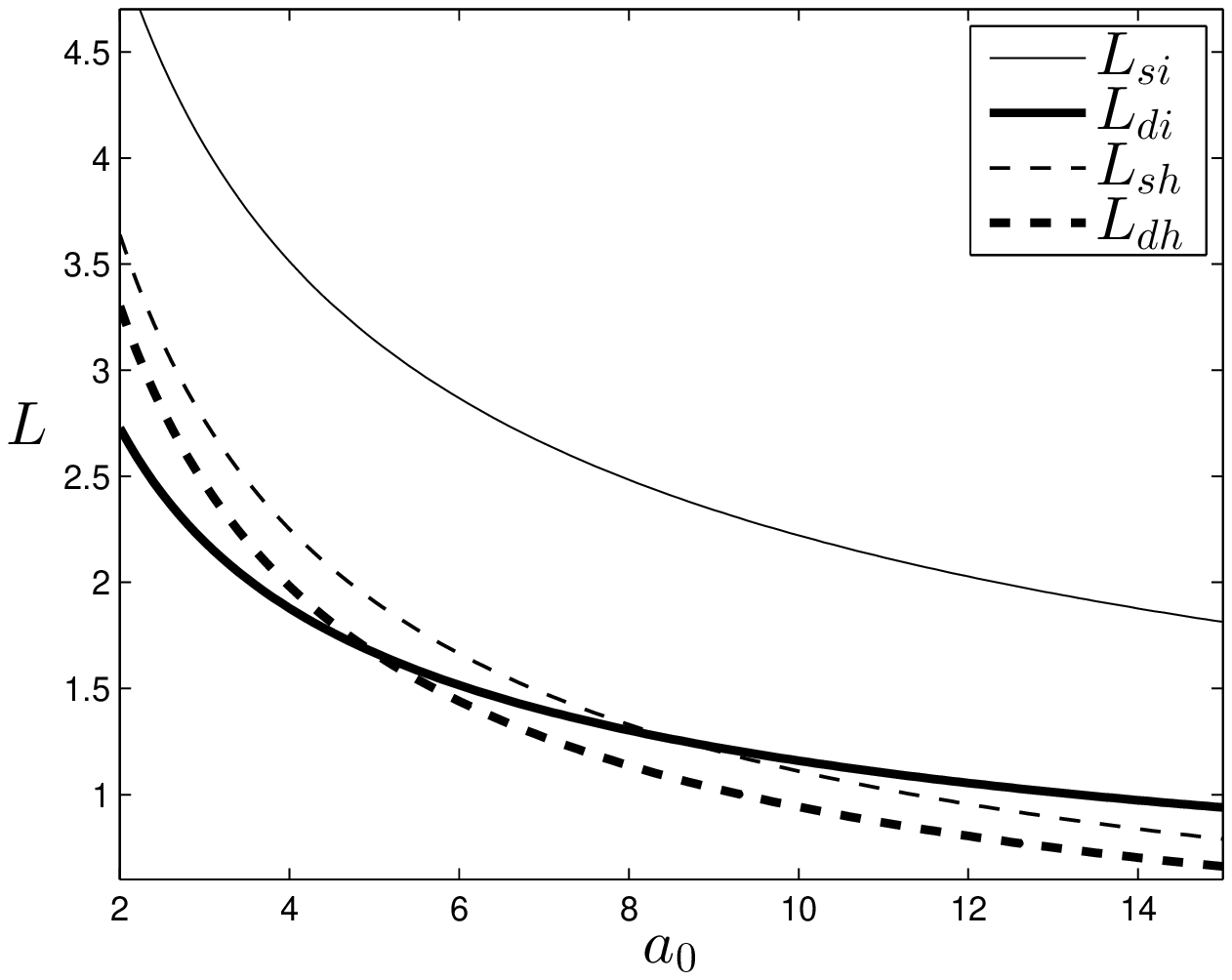}
\caption{\label{Lxa0_unexpected}  Plots of minimum size of patches versus growth rate ($a _{0}$) for systems: isolated with one ($L_{si}$) or two ($L_{di}$) patches and one ($L_{sh}$) or two ($L_{dh}$) patches immersed in the matrix for parameters $h=10$, $p=1$, $s=1$, $D=5$. Unexpected crossing of $L_{di}$ with $L_{dh}$ and $L_{sh}$ curves.}
\end{center}
\end{figure}

Now, let us summarize the conclusions of our previous work \cite{Pamplona} and the results presented by Skellam, Ludwig, Kenkre and many others \cite{Kenkre2008, Ludwig, Skellam}. It is expected that one isolated patch is the worst system to life existence, followed by a system of two communicating patches, but isolated from external matrix. When we remove the isolation condition, the worst case is the one with only one fragment, and the best case among the four cases addressed in this paper is the system with two patches immersed the matrix. In other words, the isolation is the worst factor to life existence and the second worst factor is the solitary patch, like the one presented in Fig. (\ref{Lxa0_expected}). We assume, as the worst case for life, a patch that requires a bigger size, which means $L_{dh}<L_{sh}<L_{di}<L_{si}$. These phenomenological predictions agree with the explicit expressions for the set of parameters, used in Fig. (\ref{Lxa0_expected}): $h=1$, $p=1$, $s=1$, $D=1$. This prediction do not have qualitative changes with the variation of $p$, $s$ and $D$. But if we increase the parameter $h$ while keeping the others constants, an unexpected behavior emerges from the spectrum of $a_{0}$. This behavior is presented in Fig. (\ref{Lxa0_unexpected}).

If we increase $h$, the conditions of the matrix in non isolated systems with one or two patches become worst to holding life than between the patches in the isolated system with two mutually communicating patches. Then if the conditions inside the patches, that are the same for all cases, are not very good to life, we notice that the isolated system with two communicating patches is better than non isolated systems (with one or two patches). This means that we can dribble the isolation effects by inserting a region between the patches unfavorable to life, but not fatal. This effect is related only with the increase of $h$, but it is more pronounced in systems with very diffusive populations (high values of parameter $D$), such as in Fig. (\ref{Lxa0_unexpected}), where the curves were plotted to $h=10$, $p=1$, $s=1$ and $D=5$. The dribble effect disappears for patches where the life conditions are optimum (great values of $a _{0}$). Dribbling the isolation effects with the insertion of a region unfavorable to life is not very unexpected, but the change in behavior expressed by the crossing of $L_{di}$ with $L_{dh}$ and $L_{sh}$, this is totally unexpected - see the curves in Fig. (\ref{Lxa0_unexpected}).

Increasing or decreasing the diffusion can be experimentally complicated, but if we keep the value of the life difficulty in the matrix ($h=1$) and explore the behavior of $L$ with $D$, similarly we have other unexpected relation.

We suggest an extension to the experiment proposed by Kenkre \cite{Kenkre2008} to verify our basic conclusions obtained from a simple analyze about explicit forms expressions of $L$ size. Many others conclusions can be obtained from this explicit form without a lot of effort.

\section{Conclusions}
\label{Conclusions}

The main results of this paper, namely Eq. (\ref{eqLdi}), is obtained from the explicit expression to minimum size of the non isolated system with two mutually communicating patches and its comparison with other explicit patch sizes found in the literature \cite{Kenkre2008,Ludwig,Skellam}. This comparison is possible via numerical solutions \cite{Pamplona}, but the explicit form enables us to explore the function behavior and find unexpected particularities.

The most unexpected result of this work is the change in behavior of the patch size in an isolated system with two mutually communicating patches with a fixed set of parameters and varying only the internal condition of the patches ($a_{0}$), that is the same for all cases. In Fig. (\ref{Lxa0_unexpected}) we can observe the $L_{di}$ crossing the curves $L_{sh}$ and $L_{dh}$. If this prediction is correct, for a same set of the parameters $h,p,s$ and $D$, the system studied in this work with size $L$ can be better or worst than the non isolated systems with the same size $L$, depending on just the internal conditions of the patches. Initially this is unexpected and needs experimental verification, which is left as a task to experimental researchers.

A natural continuation of this work is the investigation of a two-dimensional case, where the geometry of fragment will be a very important variable to be explored. We believe that our result can reproduce the side of a square fragment, assuming that the movement in one Cartesian direction is independent of the movement along the other one. However, we would not be surprised if the predictions of our work are found to describe the dynamics along the diameter of a circular region. If this turns out to be the case, then the movement is totally isotropic, radially symmetry. We intend to perform a thorough investigation concerning this subject in the near future.

\section*{Acknowledgments}
\label{Acknowledgments}
The authors thank PET - Programa de Educa\c{c}\~ao Tutorial for financial support and Rodrigo Rocha Cuzinatto for text revision.



\end{document}